\documentclass[epsfig,12pt] {article}
\usepackage{graphics}
\usepackage{epsfig}
\usepackage{amssymb}
\usepackage{amsmath}
\usepackage{bm}
\usepackage{epsfig}

\begin{document}

\baselineskip .7cm

\author{ Navin Khaneja \thanks{To whom correspondence may be addressed. Email:navinkhaneja@gmail.com} \thanks{Department of Electrical Engineering, IIT Bombay, Powai - 400076, India.}}

\vskip 4em

\title{\bf Chirp Excitation}

\maketitle

\vskip 3cm

\begin{center} {\bf Abstract} \end{center}
The paper describes the design of broadband chirp excitation pulses. We first develop a three stage model for understanding chirp excitation in NMR. We then show how a chirp $\pi$ pulse can be used to refocus the phase of the chirp excitation pulse. The resulting magnetization still has some phase dispersion in it. We show how a combination of two chirp $\pi$ pulses instead of one can be used to eliminate this dispersion, leaving behind a small residual phase dispersion. The excitation pulse sequence presented here allow exciting arbitrary large bandwidths without increasing the peak rf-amplitude. Experimental excitation profiles for the residual HDO signal in a sample of $99.5\%$ D$_2$O are displayed as a function of resonance offset. Although methods presented in this paper have appeared elsewhere, we present complete analytical treatment that elucidates the working of these methods.

\vskip 3em

\section{Introduction}
The excitation pulse is ubiquitous in Fourier Transform-NMR, being the starting point of all experiments. With increasing
field strengths in high resolution NMR, sensitivity and resolution comes with the challenge of
uniformly exciting larger bandwidths. At a field of 1 GHz, the target bandwidth is $50$ kHz for excitation of entire 200 ppm $^{13}$C chemical shifts. 
The required $25$ kHz hard pulse exceeds the capabilities of most
$^{13}$C  probes and poses additional problems in phasing the spectra. 
In $^{19}$F NMR, chemical shifts can range over 600 ppm, which requires excitation of different regions of the spectra. Methods that can achieve
uniform excitation over the entire bandwidth in $^{19}$F NMR, are therefore most desirable.
Towards this end, several methods have been developed for 
broadband excitation/inversion, which have reduced the phase variation of the excited magnetization as a function of the resonance offset. 
These include composite pulses, adiabatic sequences, polycromatic sequences, phase alternating
pulse sequences, optimal control pulse design, and method of multiple frames, \cite{comp1}-\cite{ultrabroadband}.

In this paper, we develop the theory of broadband chirp excitation.  We first develop a three stage model for understanding chirp excitation in NMR. We then show how a chirp $\pi$ pulse can be used to refocus the phase of the excitation pulse. The resulting magnetization still has some phase dispersion in it. We show how a combination of two chirp $\pi$ pulses instead of one can be used to eliminate this dispersion leaving behind a small residual phase dispersion. The pulse sequence presented here allow exciting arbitrary large bandwidths without increasing the peak rf-amplitude. Although methods presented in this paper have appeared elsewhere \cite{bohlen1, bohlen2, chorus}, we present complete analytical treatment that elucidates the working of these methods.

The paper is organized as follows. In section 2, we present the theory behind chirp excitation. In section 3, we present simulation and experimental results for broadband excitation pulses designed using chirp pulses. We conclude in section 4, with discussion and outlook.

\section{Theory}
Let $$ \Omega_x = \left [ \begin{array}{ccc} 0 & 0 & 0 \\ 0 & 0 & -1 \\ 0 & 1 & 0 \end{array} \right ] , \ \Omega_y = \left [ \begin{array}{ccc} 0 & 0 & 1 \\ 0 & 0 & 0 \\ -1 & 0 & 0 \end{array} \right ], \ \Omega_z = \left [ \begin{array}{ccc} 0 & -1 & 0 \\ 1 & 0 & 0 \\ 0 & 0 & 0 \end{array} \right ]. $$ 
 denote generator of rotations around $x, y, z$ axis respectively. A x-rotation by flip angle $\theta$ is 
$ \exp( \theta \Omega_x)$.

A chirp excitation pulse is understood as concatenation of three rotations

$$ \underbrace{\exp(\theta_0 \Omega_y)}_{III} \ \ \underbrace{\exp(\alpha \Omega_x)}_{II} \ \ \underbrace{\exp(\theta_0 \Omega_y)}_{I} $$

which satisfy $$ \cos \alpha = \tan^2 \theta_0. $$

Given the Bloch equation,

$$ \dot{X} = ( \omega_0 \Omega_z + A \cos \phi \ \Omega_x +  A \sin \phi \ \Omega_y ) X, $$ where $X$ is the magnetization vector,

The chirp pulse has instantaneous frequency  $ \dot{\phi} = \omega_c = - C + at$ where $a$ is the sweep rate and phase  $\phi(t) = -Ct + \frac{at^2}{2}$. The
frequency $\omega_c$ is swept from $[-C, C]$, in time $T = \frac{2C}{a}$ with offsets in range $[-B, B]$.

In the interaction frame of the chirp phase, $\phi(t)$, we have $Y(t) = \exp(-\phi(t) \Omega_x) X(t)$, evolve as  

$$ \dot{Y} = ( (\omega_0 - \omega_c) \Omega_z + A \Omega_x ) Y = \tilde \omega \ (\cos \theta(t) \ \Omega_z + \sin \theta(t) \ \Omega_y) Y,$$
where effective field strength $\tilde \omega = \sqrt{(\omega_0 - \omega_c(t))^2 + A^2}$ and $\tan \theta(t) = \frac{A}{\omega_0 - \omega_c(t)}$.

The three stages of the chirp excitation are understood in this frame.

The first rotation, $I$, arises as frequency of the chirp pulse $\omega_c$ is swept from a large negative offset $-C$ to $\omega_c - \omega_0 = - A \cot \theta_0$.  As a result, the initial magnetization follows the effective field and is transferred to 

$$ \left [ \begin{array}{c} 0 \\ 0 \\ 1 \end{array} \right ] \rightarrow \left [ \begin{array}{c} \sin \theta_0 \\ 0 \\ \cos \theta_0 \end{array} \right ]. $$

During the phase $II$ of the pulse the frequency $\omega_c - \omega_0$ is swept over the range $[- A \cot \theta_0, A \cot \theta_0]$ in time $\frac{\alpha}{A}$ and for $\cot \theta_0$ not very larger than 1, we can approximate the evolution in this phase II as 

$$ \sim \ \ \exp(\int_0^{\alpha/ A} (\omega_0 - \omega_c) \Omega_z + A \Omega_x) = \exp( \alpha \Omega_x). $$  

This produces the evolution 

$$ \left [ \begin{array}{c} \sin \theta_0 \\ 0 \\ \cos \theta_0 \end{array} \right ]  \rightarrow \left [ \begin{array}{c} \sin \theta_0 \\ -\cos \theta_0 \sin \alpha \\ \cos \theta_0 \cos \alpha \end{array} \right ]. $$

Finally, during phase $III$, the frequency is swept from $A \cot \theta_0$ to a large positive offset $C$ in time $t_f$. This produces the transformation

\begin{equation}
\label{eq:phase3}
\exp( - \underbrace{ \int \tilde \omega(t) \ dt}_{\Phi(\omega_0)}  \ \Omega_z) \exp(\theta_0 \ \Omega_y) \left [ \begin{array}{c} \sin \theta_0 \\ -\cos \theta_0 \sin \alpha \\ \cos \theta_0 \cos \alpha \end{array} \right ]. 
\end{equation}

To see this, observe, given 

$$ \dot{Y} = \tilde \omega \ (\cos \theta \ \Omega_z + \sin \theta \ \Omega_y) Y,$$

in the interaction frame of $\theta$ where $Z = \exp(- \theta(t) \Omega_y) Y $, we have, 

$$  \dot{Z}  = ( \tilde \omega \Omega_z -  \dot{\theta} \Omega_y) Z. $$

\begin{figure}
\begin{center}
\begin{tabular}{cc}
\includegraphics [scale = .3]{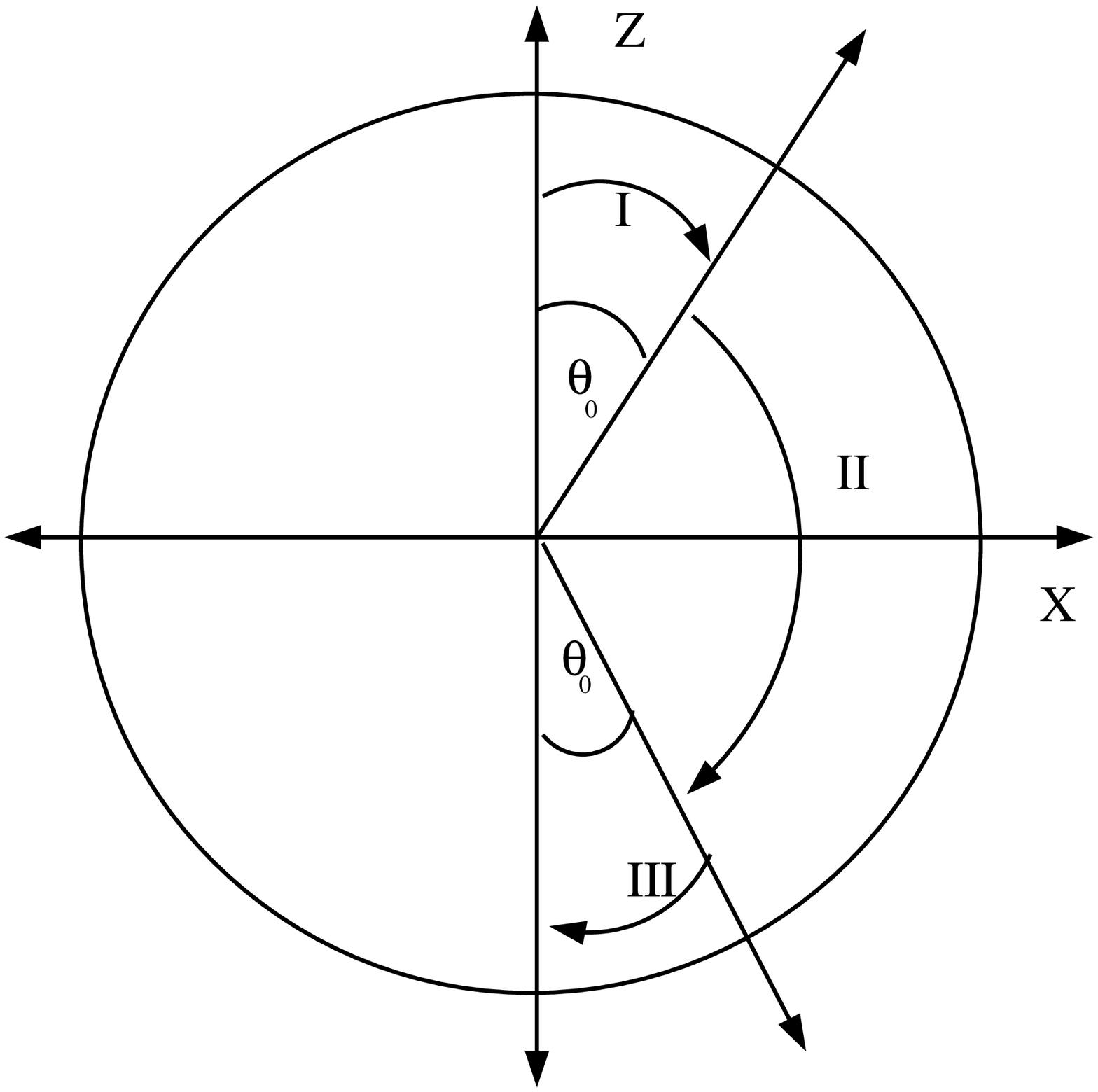} \hspace{1 in}  &
\includegraphics [scale = .3]{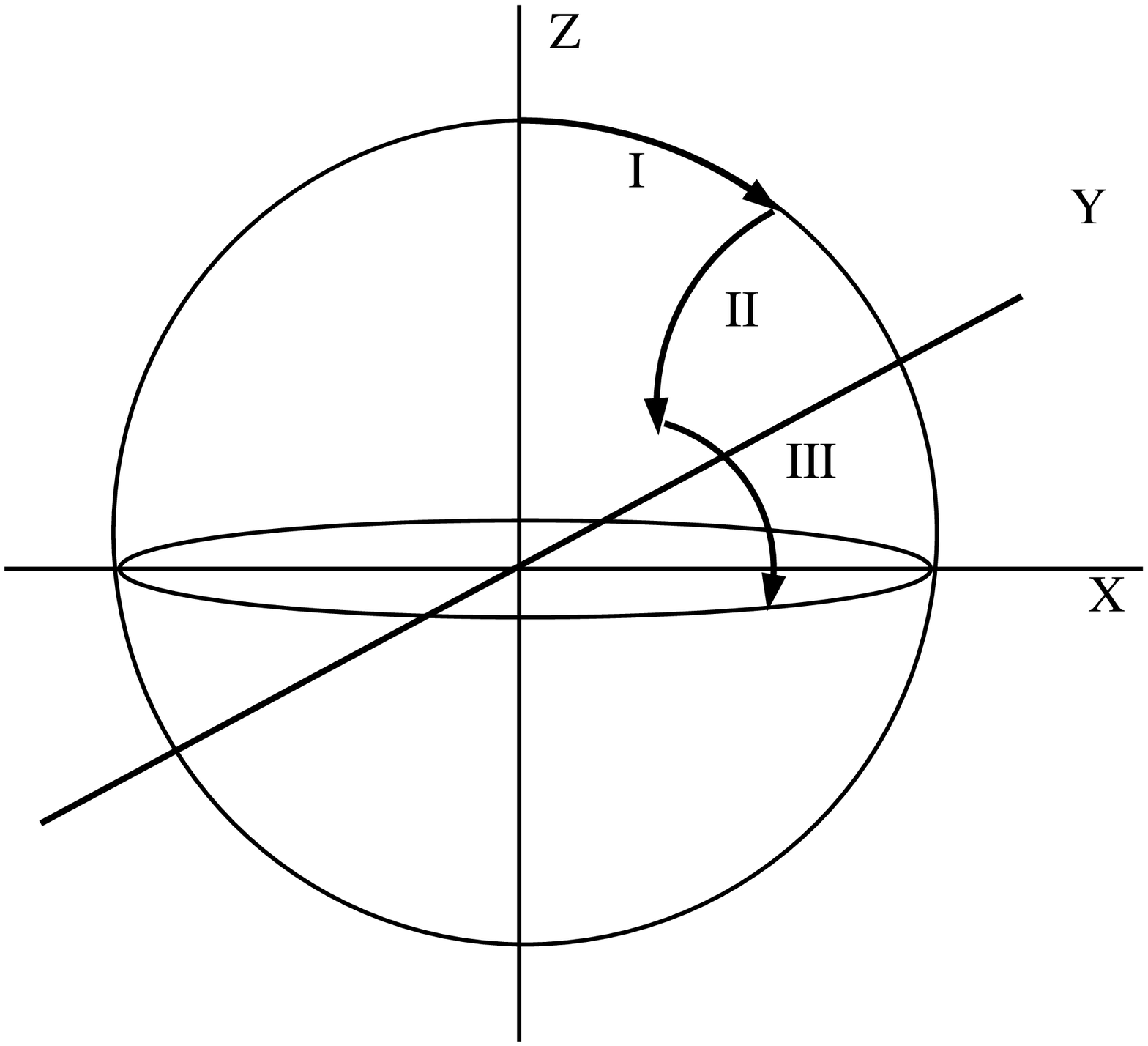}
\end{tabular}
\caption{The left panel shows the effective field for the chirp excitation. The effective field starts along $z$ axis and after phase I is rotated by $\theta_0$. After phase $II$, it makes angle of $\theta_0$ with $-z$ axis and finally at end of phase III ends up at the $-z$ axis. The right panel shows how magnetization initially along $z$ axis evolves in three stages. It 
is rotated along $y$ axis in phase I by angle $\theta_0$ and then along $x$ axis by angle $\alpha$ in phase II and finally along $y$ axis by $\theta_0$ in phase III. } \label{fig:sphere}
\end{center}
\end{figure}

If $\tilde \omega \gg \dot{\theta} $, which is true in phase III of the pulse, where $a \ll \tilde \omega^2$, as will be shown below. Then in the interaction frame of $W = \exp(- \int_0^t  \tilde \omega(t) \ \Omega_z) Z $, we average $W(t)$ to $I$. Therefore the evolution of the Bloch equation for the chirp pulse takes the form

\begin{eqnarray*}
Y(t_f) &=&  \exp( \theta(t_f) \Omega_y) Z(t_f) \\ &=&  \exp( \theta(t_f) \Omega_y) \exp( \int_0^{t_f} \tilde \omega \ \Omega_z) \underbrace{\exp(- \theta(0) \Omega_y) \ Y(0)}_{Z_0} 
\end{eqnarray*}
\begin{equation}
Y(t_f) = \exp(\pi \ \Omega_y)  \exp( \underbrace{\int \tilde \omega }_{\Phi (\omega_0)} \ \Omega_z) \exp(- \theta(0) \Omega_y ) Y(0) 
\end{equation} where $0$ marks beginning of phase III and $\theta(0) = \pi - \theta_0$, and  $\theta(t_f) = \pi$. See Fig. \ref{fig:sphere} left panel. Then this gives Eq. (\ref{eq:phase3}).

Now for this to be an excitation, the $z$ coordinate should vanish, which means,

\begin{eqnarray}
\frac{\cos \theta_0 \cos \alpha }{\sin \theta_0} &=& \tan \theta_0.  \\ 
\tan^2 \theta_0 &=& \cos \alpha.
\end{eqnarray}

For example, when $ \cot^2 \theta_0 = 2$, we have $ \cos \alpha = \frac{1}{2}$ , i.e, $\alpha = 1.0472$. Thus phase $II$ is traversed in time $\alpha A^{-1}= 1.0472 A^{-1}$. The frequency swept in this time is $2 A \cot \theta_0 = 2 A \sqrt 2$. The sweep rate is $ a = A^2 \frac{2 \sqrt 2}{1.0472} = 2.7 A^2$. The smallest effective field  in phase $I$ and $III$ is $\tilde \omega^2  = A^2(1 + \cot^2 \theta_0) = 3A^2 $ in phase $I$ and $III$. Therefore, in
  phase $I$ and $III$, we have $a \leq \tilde \omega^2 $ and adiabatic approximation is valid.
In nutshell sweep rate $a = 2.7 A^2$.

For another example, when $ \cot^2 \theta_0 = 3$, we have $ \cos \alpha = \frac{1}{3}$ , i.e, $\alpha = 1.23$. Thus phase $II$ is traversed in time $\alpha A^{-1}= 1.23 A^{-1}$. The frequency swept in this time is $2 A \cot \theta_0 = 2 A \sqrt 3$. The sweep rate is $ a = A^2 \frac{2 \sqrt 3}{1.23} = 2.81 A^2$. The smallest effective field  in phase $I$ and $III$ is $\tilde \omega^2  = A^2(1 + \cot^2 \theta_0) = 4A^2 $ in phase $I$ and $III$. Therefore, in
  phase $I$ and $III$, we have $a \leq \tilde \omega^2 $ and adiabatic approximation is valid.

In remaining paper we take $a = 2.7 A^2$. The chirp excitation doesn't produce a uniform excitation phase for all offsets.

\begin{figure}
\begin{center}
\includegraphics [scale = .5]{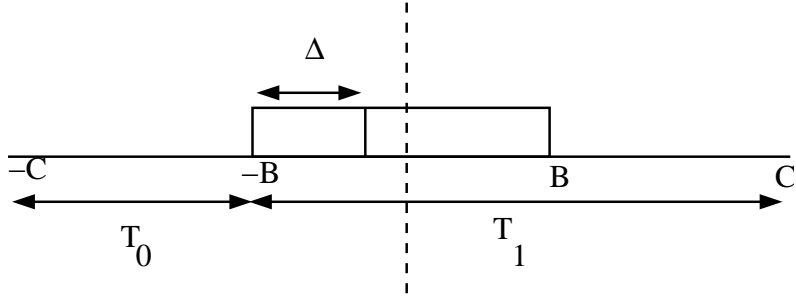}
\caption{The figure shows the offsets in range $[-B, B]$ and the sweep of chirp from $[-C, C]$. $T_1$ is the time it takes to sweep from $-B$ to $C$ and $T_0$ is the time to sweep from $-C$ to $-B$. Also shown is a offset that takes $\Delta$ time to reach from $-B$, at sweep rate $a$. } \label{fig:bandwidth}
\end{center}
\end{figure}

To understand this refer to Figure \ref{fig:bandwidth}, where offsets very from $[-B, B]$ and we sweep from $[-C, C]$ at rate $a$.
It takes $T_0$ units of time to sweep from $-C$ to $-B$ and $T_1$ units of time to sweep from $-B$ to $C$. Let $T = T_0 + T_1$ be total time. 
It takes $t_1$ units of time to sweep from $\omega_c(t) - \omega_0 = 0$ to $\omega_c(t) - \omega_0 = A \cot \theta_0$.
Then the phase $\Phi$ accumulated in Eq. \ref{eq:phase3} for the offset $-B$ is  $ \Phi(-B) = \int_{t_1}^{T_1} \tilde \omega(t) \ dt $ and for offset $-B + \Delta \omega = -B + a \Delta$
is  $ \Phi(-B + a \Delta) = \int_{t_1}^{T_1-\Delta} \tilde \omega(t) \ dt$. The difference of the phases is 

\begin{eqnarray}
\int_{T_1-\Delta}^{T_1} \tilde \omega(t) \ dt \sim  \int_{T_1-\Delta}^{T_1} (a t + \frac{A^2}{2 at}) \ dt &=& \frac{a}{2} (T_1^2 - (T_1 - \Delta)^2) + \frac{A^2}{2a} \ln \frac{T_1}{T_1 - \Delta}. \\
&=&  \frac{a}{2} (-\Delta^2 + 2 T_1 \Delta) + \frac{A^2}{2a} \ln \frac{T_1}{T_1 - \Delta}.
\end{eqnarray}

We can refocus this phase by following the chirp excitation pulse with a chirp $\pi$ pulse at twice the sweep rate $a_1 = 2 a$ and rf-field strength $A_1^2 \gg a_1$. To understand this, consider again the Bloch equation

$$ \dot{X} = ( \omega_0 \Omega_z + A_1 \cos \phi \ \Omega_x +  A_1 \sin \phi \ \Omega_y ) X, $$

where the chirp frequency  $ \dot{\phi} = \omega_c = - C + a_1 t$ is swept from $[-C, C]$.

In the interaction field of the chirp phase, $\phi(t)$, we have $Y(t) = \exp(-\phi(t) \Omega_x) X(t)$, and  

$$ \dot{Y} = ( (\omega_0 - \omega_c) \Omega_z + A_1 \Omega_x ) Y = \tilde \omega \ (\cos \theta \ \Omega_z + \sin \theta \ \Omega_y) Y,$$
where effective field strength $\tilde \omega = \sqrt{(\omega_0 - \omega_c(t))^2 + A_1^2}$ and 
$\tan \theta = \frac{A_1}{\omega_0 - \omega_c}$.

Now in interaction frame of $\theta$ where $Z = \exp(- \theta(t) \Omega_y) Y $, we have 

$$  \dot{Z}  = ( \tilde \omega \Omega_z -  \dot{\theta} \Omega_y) Z. $$

If $\tilde \omega \gg \dot{\theta} $, which is true when rf-field strength $A_1^2 \gg a_1$, in the interaction frame of $W = \exp(- \int_0^t  \tilde \omega \ \Omega_z) Z $, we average $W(t)$ to $I$. Therefore the evolution of the Bloch equation for the chirp pulse takes the form

\begin{equation}
X(t) = \exp(\phi(t) \ \Omega_z) \exp(\theta(t) \ \Omega_y)  \exp( \underbrace{\int_0^t \tilde \omega }_{\Phi_1 (\omega_0)} \ \Omega_z) X(0) 
\end{equation}

where $\phi(0) = \phi(T) = 0$ and $\theta(0) = 0$ and $\theta(T) = \pi$ and now we can again evaluate 
$\Phi_1(-B)- \Phi_1(-B + a \Delta) $. Observe

\begin{eqnarray*}
\Phi_1(-B) &=& \int_0^{\frac{T_0}{2}} \sqrt{(a_1 t)^2 + A_1^2} \ dt + \int_0^{\frac{T_1}{2}} \sqrt{(a_1 t)^2 + A_1^2} \ dt.  \\
 \Phi_1(-B + a \Delta) &=& \int_0^{\frac{T_0+ \Delta}{2}} \sqrt{(a_1 t)^2 + A_1^2} \ dt + \int_0^{\frac{T_1-\Delta}{2}} \sqrt{(a_1 t)^2 + A_1^2} \ dt. \\
\Phi_1(-B)- \Phi_1(-B + a \Delta)  &=& \int_{\frac{T_1-\Delta}{2}}^{\frac{T_1}{2}} \sqrt{(a_1 t)^2 + A_1^2} \ dt - \int_{\frac{T_0}{2}}^{\frac{T_0+ \Delta}{2}} \sqrt{(a_1 t)^2 + A_1^2}\ dt . \\
 \int_{\frac{T_1-\Delta}{2}}^{\frac{T_1}{2}} \sqrt{(a_1 t)^2 + A_1^2} \ dt &\sim & \frac{a}{4} (T_1^2 - (T_1 - \Delta)^2) + \frac{A_1^2}{4a} \ln \frac{T_1}{T_1 - \Delta}. \\
\Phi_1(-B)- \Phi_1(-B + a \Delta) &\sim& \frac{a}{4} (-2\Delta^2 + 2 (T_1 -T_0) \Delta )  +  \frac{A_1^2}{4a} \ln \frac{T_1T_0}{(T_1 - \Delta)(T_0 + \Delta)}.
\end{eqnarray*}

Now if we combine the phase due to chirp excitation excitation pulse and the chirp $\pi$ pulse we get 

\begin{eqnarray*}
 \{ \Phi_1(-B + a \Delta) -  \Phi_1(-B) \} - \{ \Phi(-B + a \Delta) - \Phi(-B) \} &=& \frac{a (T_1 + T_0) \Delta}{2}  -   \frac{A_1^2}{4a} \ln \frac{T_1T_0}{(T_1 - \Delta)(T_0 + \Delta)} \\ &+&  \frac{A^2}{2a} \ln \frac{T_1}{T_1 - \Delta}.
\end{eqnarray*}

\begin{figure}
\begin{center}
\includegraphics [scale = .5]{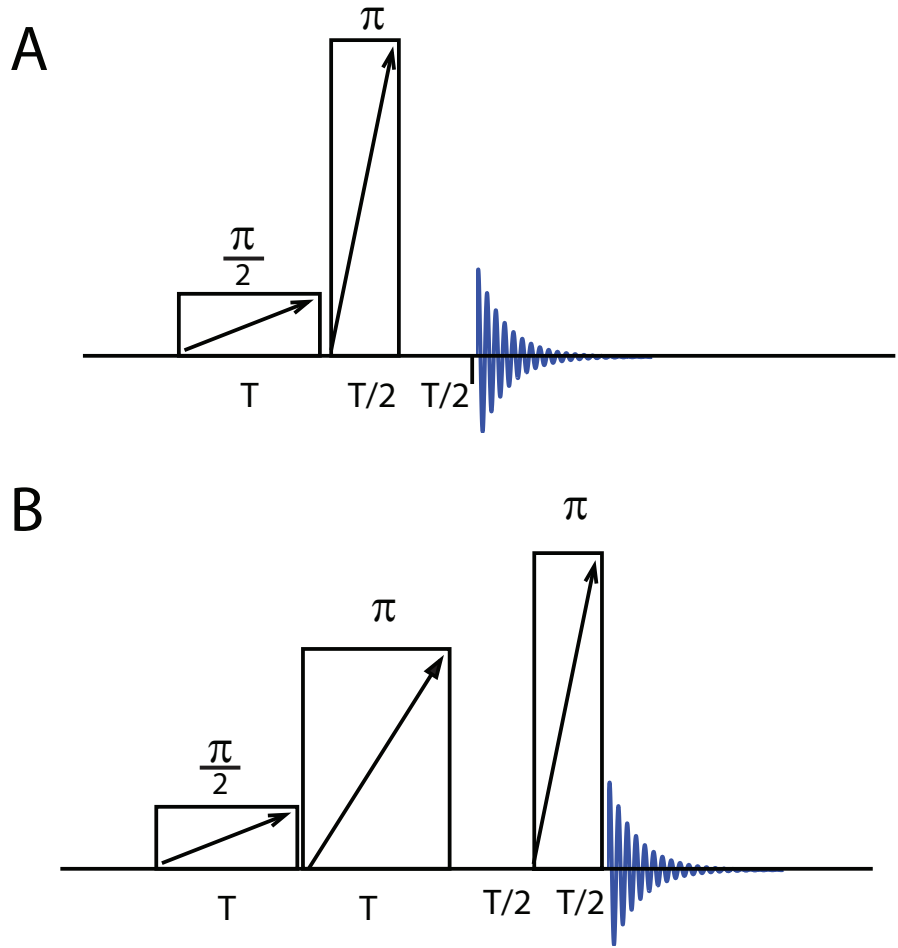}
\caption{Fig. A shows the pulse sequence with a $\frac{\pi}{2}$ excitation pulse of duration $T$ followed by a $\pi$ inversion pulse of duration $\frac{T}{2}$ at twice the sweep rate and finally a free evolution for time $\frac{T}{2}$. Fig. B shows the pulse sequence with a $\frac{\pi}{2}$ excitation pulse of duration $T$ followed by a $\pi$ inversion pulse of duration $T$ both at same sweep rate and finally a free evolution for time $\frac{T}{2}$ followed by a $\pi$ inversion pulse of duration $\frac{T}{2}$ at twice the chirp rate. The ratio of amplitude of last $\pi$ pulse to center $\pi$ pulse is $\sqrt{2}$.} \label{fig:2pulse}
\end{center}
\end{figure}

If chirp $\pi$ pulse is followed by free evolution for $\frac{T}{2}$ where $T = T_1 + T_0$, it refocuses the phase $\frac{a (T_1 + T_0) \Delta}{2} = \frac{T \Delta \omega}{2}$. See Fig. \ref{fig:2pulse}A. The only phase dispersion that is left is 

\begin{equation}
\label{eq:totaldispersion}
\frac{A_1^2}{4a} \ln \frac{(T_1 - \Delta)(T_0 + \Delta)}{T_1T_0} +  \frac{A^2}{2a} \ln \frac{T_1}{T_1 - \Delta} 
\end{equation}

For $a \Delta = 2 B$, the other extreme of the spectrum, the above expression simplifies to 

\begin{equation}
\label{eq:phasedisp}
\frac{A^2}{2a} \ln \frac{1 + \frac{B}{C}}{1 - \frac{B}{C}}. 
\end{equation}
As described before for $ a = 2.7 A^2$ and when $\frac{B}{C} \ll 1 $ say $\frac{B}{C} = 1/3$, this dispersion is small around $7^{\circ}$.

The factor $\frac{A_1^2}{4a} \ln \frac{(T_1 - \Delta)(T_0 + \Delta)}{T_1T_0}$ in Eq. (\ref{eq:totaldispersion}) can be cancelled by introducing a $\pi$ pulse of 
amplitude $\frac{A_1}{\sqrt{2}}$, and sweep rate $a$, following $\frac{\pi}{2}$ chirp pulse and then a delay of $\frac{T}{2}$, and finally
the  $\pi$ pulse of  amplitude $A_1$ and sweep rate $2a$. See Fig. \ref{fig:2pulse}B. Then all phase dispersion cancel except 
the one in Eq. (\ref{eq:phasedisp}). We can make this dispersion small by  $\frac{B}{C} \ll 1 $.

\section{Simulation and Experiments}

\begin{figure}
\begin{center}
\includegraphics [scale =.5]{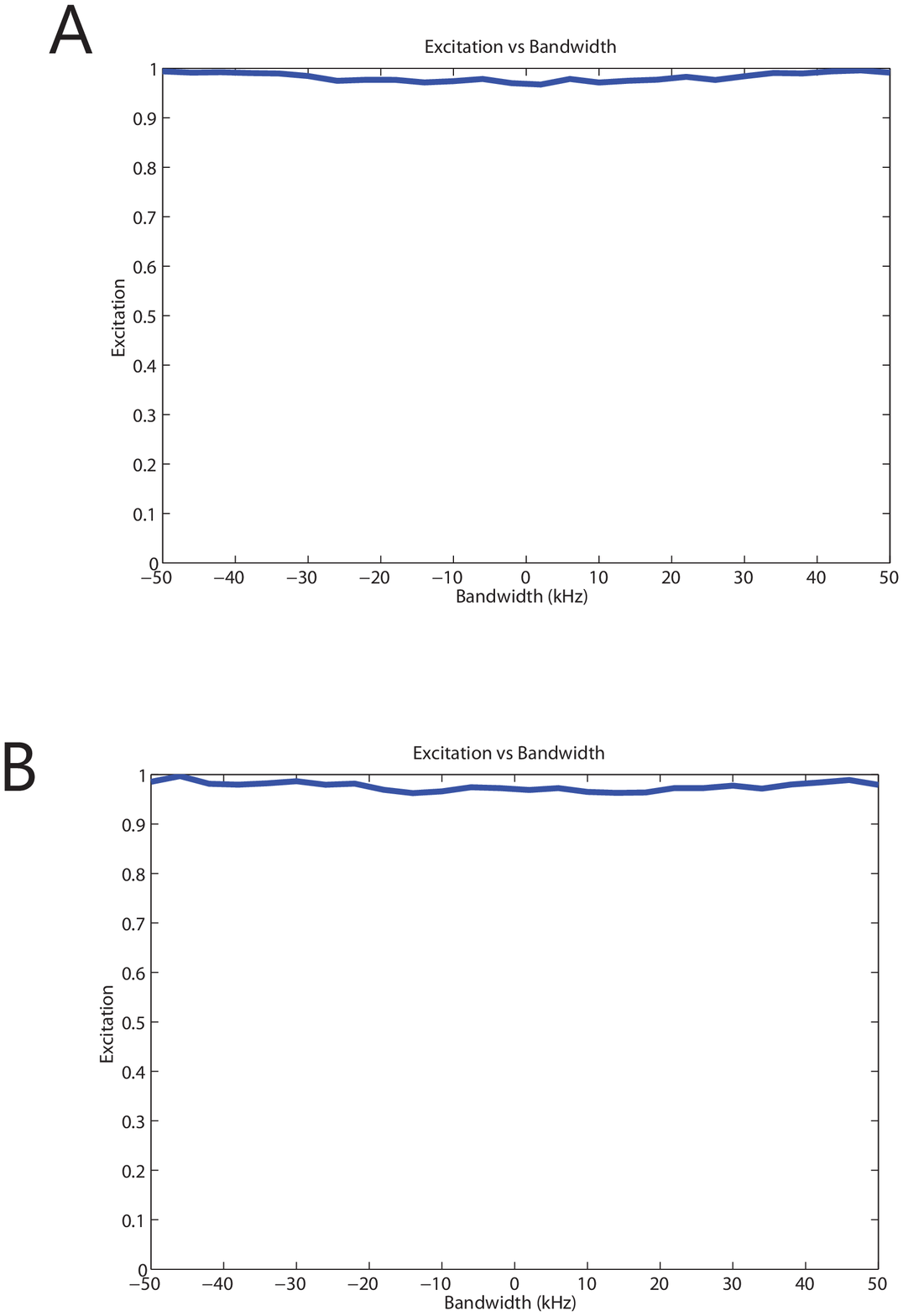}
\caption{In Fig. A we choose amplitude of $\pi$ pulse as $10$ kHz and $\frac{\pi}{2}$ pulse as $1$ kHz. The Bandwidth
$B/2 \pi = 50$ kHz and $C/2 \pi = 400$ kHz. Sweep rate $a = 2.7 \times (2 \pi \ kHz)^2$ and time $T = 47.15$ ms. Total duration of the pulse is 
$94.13$ ms. In Fig. B we choose amplitude of last $\pi$ pulse as $10$ kHz, center $\pi$ pulse as $\frac{10}{\sqrt{2}}$ kHz and $\frac{\pi}{2}$ pulse as $1$ kHz. The Bandwidth
$B/2 \pi = 50$ kHz and $C/2 \pi = 150$ kHz. Sweep rate $a = 2.7 \times (2 \pi \ kHz)^2$ and time $T = 17.68$ ms. Total duration of the pulse is 
$53.0516$ ms. } \label{fig:2pls3pls}
\end{center}
\end{figure}

In Fig. \ref{fig:2pulse}A, we choose amplitude of $\pi$ pulse as $10$ kHz and $\frac{\pi}{2}$ pulse as $1$ kHz. The bandwidth
$B/2\pi = 50$ kHz and $C/2 \pi = 400$ kHz. Sweep rate $a = (2 \pi \ kHz)^2$ and time $T = 47.15$ ms. Total duration of the pulse is 
$94.13$ ms. Fig. \ref{fig:2pls3pls}A, shows the x-coordinate of the excited magnetization, after a zero order phase correction.

In Fig. \ref{fig:2pulse}B, we choose amplitude of last $\pi$ pulse as $10$ kHz, center $\pi$ pulse as $\frac{10}{\sqrt{2}}$ kHz and $\frac{\pi}{2}$ pulse as $1$ kHz. The bandwidth
$B/2 \pi = 50$ kHz and $C/2 \pi = 150$ kHz. Sweep rate $a = (2 \pi \ kHz)^2$ and time $T = 17.68$ ms. Total duration of the pulse is 
$53.0516$ ms. Fig. \ref{fig:2pls3pls}B, shows the x-coordinate of the excited magnetization, after a zero order phase correction.

\begin{figure}
\begin{center}
\includegraphics [scale =.4]{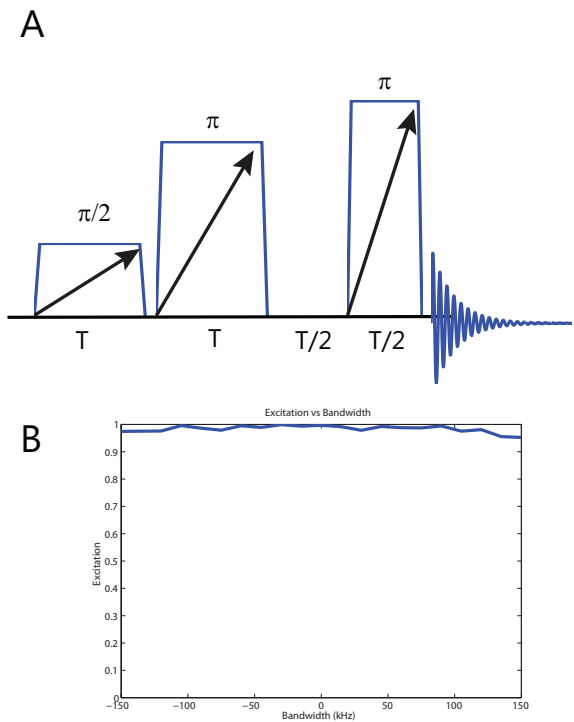}
\caption{Fig. A shows the pulse sequence with a $\frac{\pi}{2}$ excitation pulse of duration $T$ followed by a $\pi$ inversion pulse of duration $T$ both at same sweep rate and finally a free evolution for time $\frac{T}{2}$ followed by a $\pi$ inversion pulse of duration $\frac{T}{2}$ at twice the chirp rate. The amplitude at ends of chirp pulses is tapered to minimize sweep width. The ratio of peak amplitude of last $\pi$ pulse to center $\pi$ pulse is $\sqrt{2}$. Fig. B shows the excitation profile after a zero order phase correction of the pulse sequence in Fig. A. We choose peak amplitude of last $\pi$ pulse as $15$ kHz, center $\pi$ pulse as $\frac{15}{\sqrt{2}}$ kHz and $\frac{\pi}{2}$ pulse as $3$ kHz. The bandwidth
$B/2 \pi = 150$ kHz and $C/2 \pi = 180 $ kHz. Sweep rate $a = 2.7 \times (2 \pi \times 3 \ kHz)^2$ and time $T = 2.36$ ms. Total duration of the pulse sequence is $7.07$ ms. } \label{fig:garret}
\end{center}
\end{figure}

Finally, in Fig. \ref{fig:garret}A, we taper the edges of chirp pulse so that we don't have to sweep very far. We choose peak Amplitude of last $\pi$ pulse as $15$ kHz, center $\pi$ pulse as $\frac{15}{\sqrt{2}}$ kHz and $\frac{\pi}{2}$ pulse as $3$ kHz. The Bandwidth
$B/2 \pi = 150$ kHz and $C/2 \pi = 180 $ kHz. Sweep rate $a = 2.7 \times (2 \pi \times 3 \ kHz)^2$ and time $T = 2.36$ ms. Total duration of the pulse sequence is $7.07$ ms. Fig. \ref{fig:garret}B, shows the x-coordinate of the excited magnetization, after a zero order phase correction.

\begin{figure}
\begin{center}
\includegraphics [scale =.7]{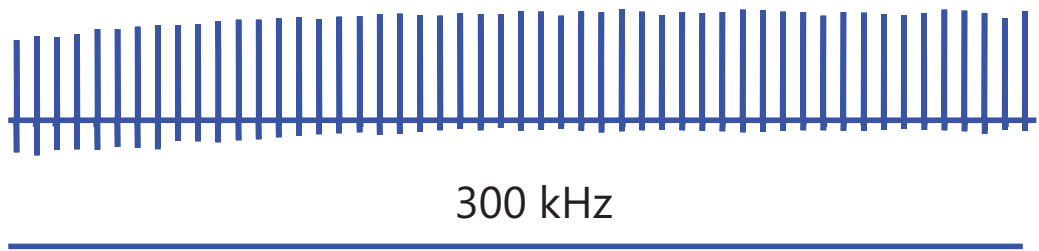}
\caption{Above figure shows experimental excitation profiles of residual HDO signal in a sample of $99.5\%$ D$_2$O, as a function of resonance offset, after application of the pulse sequence in Fig. \ref{fig:garret}A. The offset is varied in increments of 6 kHz, over a range of 300 kHz as in fig. \ref{fig:garret}B} \label{fig:chirp}
\end{center}
\end{figure}

Experimental realization of this pulse sequence is done on a 750 MHz spectrometer. Fig. \ref{fig:chirp} shows experimental excitation profiles of residual HDO signal in a sample of $99.5\%$ D$_2$O, as a function of resonance offset, after application of the pulse sequence in Fig. \ref{fig:garret}A. The offset is varied in increments of 6 kHz, over a range of 300 kHz ([ -150, 150] kHz) around the proton resonance at 4.7 ppm. The peak amplitude of the pulse is 15 kHz. We find uniform excitation experimentally.

This pulse sequence appears as CHORUS in \cite{chorus}. Here we provide details of the working of this sequence.

\section{Conclusion}
In this paper we developed the theory of broadband chirp excitation pulses. We first developed a three stage model for understanding chirp excitation in NMR. We then showed how a chirp $\pi$ pulse can be used to refocus the phase of the chirp excitation pulse. The resulting magnetization still had some phase dispersion in it. We then showed how a combination of two chirp $\pi$ pulses instead of one can be used to eliminate this dispersion, leaving behind a small residual phase dispersion. The excitation pulse sequence presented here allow exciting arbitrary large bandwidths without increasing the peak rf-amplitude. They are expected to find immediate application in 
 $^{19}$F NMR. Although methods presented in this paper have appeared elsewhere \cite{bohlen1, bohlen2, chorus}, we present complete analytical treatment that elucidates the working of these methods. Future work in direction is to use these methods to develop
general purpose rotation pulses, like a broadband $(\frac{\pi}{2})_x$ rotation.

\end{document}